\documentclass[12pt]{article}
\usepackage{epsf,latexsym}
\usepackage[all]{xy}
\usepackage{amsfonts,amssymb} 
\usepackage{hyperref}
\epsfverbosetrue
\newcommand{\de}{\hbox{\rm{d}}}

\newcommand{\bb}{\begin{eqnarray}}
\newcommand{\ee}{\end{eqnarray}}
\newcommand{\eee}{\nonumber\end{eqnarray}}
\newcommand{\qq}{\quad}

\begin{document}

\font\twelve=cmbx10 at 13pt
\font\eightrm=cmr8

\thispagestyle{empty}

\begin{center}
${}$
\vspace{3cm}

{\Large\textbf{Lensing in an interior Kottler solution}} \\

\vspace{2cm}

{\large Thomas Sch\"ucker\footnote{also at Universit\'e de Provence, Marseille,
France, thomas.schucker@gmail.com } (CPT\footnote{Centre de Physique
Th\'eorique\\\indent${}$\qq\qq CNRS--Luminy, Case
907\\\indent${}$\qq\qq 13288 Marseille Cedex 9,
France\\\indent${}$\qq
Unit\'e Mixte de Recherche (UMR 6207) du CNRS et des Universit\'es
Aix--Marseille 1 et 2\\
\indent${}$\qq et Sud Toulon--Var, Laboratoire affili\'e \`a la
FRUMAM (FR 2291)})}

\vspace{2cm}

\vskip 2cm

{\large\textbf{Abstract}}
\end{center}
We derive the interior Kottler solution of the incompressible fluid and show that the bending of light in this solution does depend on the cosmological constant.
\vspace{2cm}

\vskip 2truecm

\noindent CPT-P004-2009 \\
\noindent 0903.2940
\vspace{1cm}

\section{Introduction}

It is well known that in Kottler's (or Schwarzschild-de Sitter's) metric the spatial part of the geodesic $\varphi (r)$ of a massless particle in static polar coordinates does not depend on the cosmological constant $\Lambda $. From this it was concluded that physical lensing angles also were independent of $\Lambda $. Only recently,
Rindler \& Ishak \cite{ri} corrected this widely held believe. They point out that $\Lambda $ re-enters the scene when coordinate angles are translated into physical angles. However the community has not accepted unanimously that the bending of light does change with the cosmological constant. References in \cite{pro} support the claim, those in \cite{contra} refute it, while reference \cite{syn} proposes a synthesis. Ishak, Rindler et al. \cite{ir} go one step further and look for lens systems where the $\Lambda $ dependence is observable. In these systems the lens is a galaxy cluster with an extension larger than the pericenter of the light rays. It is therefore interesting both with respect to the open controversy as well as for practical purposes to analyse the $\Lambda $ dependence of lensing in an interior Kottler solution. For simplicity, we will use the one of the incompressible fluid here.

\section{Kottler's solution for the incompressible fluid}

The most general static, spherical metric is of the form:
\bb \de \tau^2=B\,\de t^2-A\, \de r^2-r^2\de\theta ^2-r^2\sin ^2\theta \,\de\varphi ^2,\label{stat}\ee
where $A$ and $B$ are functions of $r$. Consider a ball of constant mass density $\rho $ centered in $r=0$, of total mass $M$, and of radius $R$,
$ \rho =M/({\textstyle\frac{4}{3}}\pi R^3)$ .
The exterior Kottler solution,
\bb A=\,\frac{1}{B}\,,\qq B=1-\,\frac{2GM}{r}\, -{\textstyle\frac{1}{3}}  \Lambda r^2,\qq r\ge R,\ee
has vanishing pressure $p$.
The interior solution, $r\le R$, is the unique solution of Einstein's equations,
\bb \,\frac{1}{r^2}\, -\,\frac{1}{r^2}\,\left( \,\frac{r}{A}\, \right) '-\Lambda &=&8\pi G\,\rho ,\qq\qq':=\,\frac{\de}{\de r}\, ,\\
-\,\frac{1}{r^2}\, +\,\frac{1}{r^2A}\, +\,\frac{1}{rA}\,\frac{B'}{B}+\Lambda&=&8\pi G \,p\, ,\ee
(and of the continuity equation,
$p'+{\textstyle\frac{1}{2}} ({B'}/{B})\,(\rho +p)=0\,$)
with $A(r)$, $B(r)$ and $p(r)$ continuous at the boundary $r=R$. These equations have seperate variables and can be integrated readily. The solution is conveniently written in terms of auxiliary quantities:
\bb \gamma :={\textstyle\frac{1}{3}} (8\pi  G\,\rho +\Lambda ),&\alpha :={{\textstyle\frac{1}{2}}
 8\pi G\,\rho }/{\gamma } ,&
\beta :=\,\frac{-{\textstyle\frac{1}{6}} 8\pi G\,\rho\,+\,{\textstyle\frac{1}{3}} \Lambda }{\gamma }\,=\,1-\alpha  ,\\
w(r):=\sqrt{1-\gamma r^2},&
K:=w(R),&\\
A=\,\frac{1}{w^2}\, ,& \qq B=(\alpha K+\beta w)^2,\qq&
p=\rho \,\left[ \,\frac{K}{\alpha K+\beta w}\, -1\right] .\ee
This solution agrees with the ones by Stuchl\'{i}k \cite{st} and Boehmer \cite{bo}.
For vanishing cosmological constant, $\alpha =3/2$ and $\beta =-1/2$, we retrieve Schwarzschild's famous interior solution. Note that simply substituting $\rho +\Lambda /(8\pi G)$ for the density and $p -\Lambda /(8\pi G)$ for the pressure in 
Schwarzschild's solution does yield a solution of Einstein's equations with cosmological constant, but with discontinuous pressure.

\section{ Geodesics of massless particles}

Because of spherical symmetry the photons travel in a plane containing the origin. We take this plane to be $\theta =\pi /2$. Then the spatial part of the geodesic of a massless particle in the static spherical metric (\ref{stat}) reads:
\bb \left( r \,\frac{\de \varphi }{\de r}\, \right) ^{-2}=
\,\frac{1}{A(r)}\, \left( \,\frac{r^2}{R^2} \,\frac{C+\,B(R)}{B(r)}\, -1\right) ,\label{spatial}\ee
with initial condition $C:=(R\,\de \varphi (R) /\de r)^{-2}$. 
It is a remarkable fact \cite{gg} that the spatial form -- in these coordinates --  of any geodesic of a massless particle, i.e.  the right-hand side of equation (\ref{spatial}),  is strictly independent of $\Lambda $ for the exterior Kottler solution. In particular, this is not true for geodesics of a massive particles.

It is straight forward to check that the geodesic of a massless particle in the interior Kottler solution for the incompressible fluid does depend on the cosmological constant. However, this dependence is tiny even for big clusters, $M= 10^{15} {\rm M}_\odot,$ $R=3\cdot10^{23}$ m, and a realistic cosmological constant, $\Lambda =1.5\cdot 10^{-52}\ {\rm m}^{-2}\ \pm 20\ \%.$ Indeed then $8\pi G\,\rho =5.8\cdot10^{-51}\ {\rm m}^{-2},$ $8\pi G\,\rho R^2 =5.2\cdot10^{-4},$ and $\Lambda R^2=1.4\cdot 10^{-5}.$ Developing the right-hand side of equation (\ref{spatial}) we get:
\bb \left( r \,\frac{\de \varphi }{\de r}\, \right) ^{-2}&=&-1+(C+1)\,\frac{r^2}{R^2}\, 
+{\textstyle\frac{1}{2}} (C+1)\, \left( 1-\,\frac{r^2}{R^2}\right) \,( 8\pi G\rho r^2)\,\nonumber\\[2mm]
&&+{\textstyle\frac{1}{8}} \left[ -{\textstyle\frac{7}{3}} C-1+{\textstyle\frac{1}{2}}(C+1) \,\frac{r^2}{R^2}\,+{\textstyle\frac{1}{2}}({\textstyle\frac{11}{3}} C+1) \,\frac{R^2}{r^2}\,\right] \,( 8\pi G\rho r^2)^2\nonumber\\[2mm]
&&+{\textstyle\frac{1}{4}}\left[ 
{\textstyle\frac{1}{6}} (C-3) 
+{\textstyle\frac{1}{3}}(C+3) \,\frac{r^2}{R^2}\,
- {\textstyle\frac{1}{2}}(C+1) \,\frac{r^4}{R^4}\, \right]\, (8\pi G\rho r^2)\,(\Lambda R^2) \nonumber\\[2mm]&& +\ {\rm higher \ order\ terms.}\label{perturb}
\ee 
The leading term in $\Lambda$ is suppressed by one power of $8\pi G\,\rho R^2 =5.2\cdot10^{-4}$. Its sign is such that a positive cosmological constant may enhance the bending of light. 

As noted by Rindler \& Ishak \cite{ri},  a static observer at the outside of the fluid will observe an angle that is different from the coordinate angle $\varphi $ by a factor $\sqrt{1-2GM/r_{\rm obs}-\Lambda r_{\rm obs}^2/3} $ where $r_{\rm obs}$ is the coordinate distance of the observer from the center. This factor weakens the bending of light. I have no idea why the two dependencies, in the interior and in the exterior, are of opposite sign, but the latter certainly prevails for observers far outside the lens. 

\begin{center}
\begin{tabular}{c}

\xy
(0,0)*{}="L";
(-50.25,0)*{}="S";
(50,5)*{}="E";
(-50.25,0)*{\bullet};
(50,5)*{\bullet};
(-54.3,0)*{S};
(54,6)*{O};
(-2.3,-2.9)*{L};
(0,0)*{\bullet};
{\ar (0,0)*{}; (55,0)*{}}; 
"L"; "E" **\dir{-}; 
"L"; (-1,20) **\dir{-}; 
"E"; (-1,20) **\dir{-};
(-1,20); (20,28.5) **\dir{-};
"S"; (-1,20) **\dir{-};
(4.5,22.23);(4.5,18.23) **\crv{(6,20)};
(8,20.5)*{\delta  };
(-45,2.13);(-44.5,0) **\crv{(-43.8,1.1)};
(-41,1.5)*{\epsilon };
(44,6.6);(44,4.4) **\crv{(43,5.5)};
(40.7,5.6)*{\epsilon };
(5,0);(-0.25,5) **\crv{(4.4,4.4)};
(6,5)*{\varphi _P};
(-24,-2.5)*{R};
(23.6,5)*{R};
(60,0)*{x};
(0,-13)*{};
"S"; "L" **\dir{-}; 
"E"; "S" **\crv{(-1,20)};
\endxy
\end{tabular}\linebreak\nopagebreak
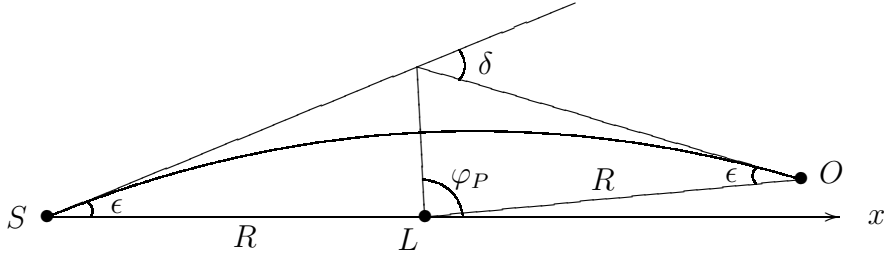
{Figure 1: The scattering angle $\delta $}
\end{center}

Let us give an example. Suppose that at the coordinate distance $R$, a source $S$ emits a light ray  under a coordinate angle $\epsilon =5^\circ$ with respect to the direction towards the center $L$ of the lens, $\tan \epsilon = R\,\de \varphi (R) /\de r$. The out-going light ray is observed again at coordinate distance $R$, see figure 1. Let $\delta = \pi +2\,\epsilon-2\,\varphi (r_P) $ be the scattering angle, where
 $r_P$ is the coordinate value of the pericenter. 
 Likewise $\delta $ is a coordinate angle.
 In its perturbative approximation, the integral of equation (\ref{perturb}) yields an elliptic function. For reasons explained below, the integral is done numerically. Due to the integrable singularity at the pericenter $r_P$, this evaluation is delicate. We take $M= 10^{15} {\rm M}_\odot,$ $R=3\cdot10^{23}$ m, and two values for the  cosmological constant, $\Lambda =0$ and $\Lambda =1.5\cdot 10^{-52}\ {\rm m}^{-2}.$  We find:
$r_{P0} = 2.6\cdot10^{22}$ m, $\delta _0 = 0.3''$ and 
$r_{P0}-r_P = 1.8\cdot10^{11}$ m, $\delta = 0.5''$.

Of course we must still translate these coordinate results using physical lengths and angles. Let us imagine a standard candle at the center, $r=0$, and let us use it to measure the luminosity distance of the source $S$ of our light ray and of the observer $O$  from the center. These luminosity distances both coincide with the coordinate distance $R$.  Notice that we cannot take a standard candle as emitter of our light ray because then - in the case of strong lensing with aligned source, lens and observer -  the observer would receive an infinite apparent luminosity and conclude that her distance from the source is zero.

 The relation between the coordinate angle $\epsilon$ and the physical angle at distance $R$ is given by a factor $\sqrt {B(R)}\sim1-3\cdot 10^{-4}$, which can safely be neglected in our situation. In our example the light bending is too weak to produce a double image.  Therefore the physical measurement of the scattering angle would require a third star far away from the lens and a triangularization with $S$ and $O$, which is of course academic.

Equation (\ref{perturb}) tells us that the interior contribution of the cosmological constant to
the
deflection is $GM\Lambda$ times a characteristic length scale. This can be compared qualitatively with a result by  Sereno  in \cite{se}. He discusses light propagation
in the exterior Kottler metric including higher order terms and tries
to extrapolate that result to
estimate contributions for light rays propagating inside galaxy
clusters.  He finds an additional contribution  proportional to $GM\Lambda b$
 with $b$ being the impact parameter, $b\sim \epsilon r_P$ for small $\epsilon$. The pericenter $r_P$ is  a root of the right hand side of equation (\ref{perturb}). Therefore the above length scale depends on the density profile $\rho (r)$ in the interior. For constant $\rho $, a numerical example of the pericenter is given above.   As Sereno's extrapolation is independent of $\rho $, a quantitative comparison is not attempted.

\section{Conclusion}

Our calculation shows by an explicit example that the differential equation for light rays in an interior Kottler solution does depend on
the cosmological constant, in contrast to the exterior
(vacuum) Kottler solution where the cosmological constant
influences the light bending only indirectly through the translation of coordinate angles into physically observed angles.

[It is sometimes claimed that the cosmological constant only influences the bending of light via its influence on the luminosity distances. The present computation shows that this claim is wrong in our particular situation. What is more, the general claim is ill defined because of magnification, except in a Robertson-Walker spacetime, where indeed the cosmological constant changes luminosity distances and where magnification is absent. But there, lensing is absent as well. {\it One of the three anonymous referees had me suppress this passage.}]

The present computation also tells us that it is not easy to find a concrete physical system where the influence of the cosmological constant on the lensing angles is accessible experimentally. Of course the constant mass density, that we have considered to simplify the computations, is unrealistic and we are still working on numerical calculations with more realistic density profiles, like the one by Navarro, Frenk \& White \cite{nfw}.

The simple question whether or not the bending of light by a single spherical mass does depend on the cosmological constant has far reaching conceptual consequences, if the answer is affirmative. Indeed, then two less innocent questions pop up immediately:
\begin{itemize}\item
How can we generalize the analysis of multiple scattering to include a cosmological constant? This question is difficult because the repulsive force induced by a positive $\Lambda $ outside the mass distribution is long range, it even increases with distance. The interior Kottler solution is thought to describe a (highly symmetric) distribution of scattering centers. In this sense, the present result indicates that the cosmological constant does play a role already inside the matter.
\item
How do fashionable generalisations of the cosmological constant like dark matter, quintessence, ... modify the bending of light by a single spherical mass? 
\end{itemize}

\end{document}